\documentclass[prl,twocolumn]{revtex4-2}
\usepackage{amsmath}
\usepackage{amssymb}
\usepackage{graphicx}
\usepackage{braket}
\usepackage{stmaryrd}
\usepackage{hyperref}
\usepackage{color}
\usepackage{dsfont}

\hypersetup{
    colorlinks,
    citecolor=blue,
    linkcolor=blue,
    urlcolor=blue
}

\begin{document}
\title{Quantum Thermodynamic Uncertainty Relation under Feedback Control}
\author{Yoshihiko Hasegawa}
\email{hasegawa@biom.t.u-tokyo.ac.jp}
\affiliation{Department of Information and Communication Engineering, Graduate
School of Information Science and Technology, The University of Tokyo,
Tokyo 113-8656, Japan}
\date{\today}
\begin{abstract}
The thermodynamic uncertainty relation posits that higher thermodynamic costs are essential for a system to function with greater precision. 
Recent discussions have expanded thermodynamic uncertainty relations beyond classical non-equilibrium systems, investigating how quantum characteristics can be utilized to improve precision. 
In this Letter, we explore how quantum feedback, a control technique used to manipulate quantum systems, can enhance the precision. 
Specifically, we derive a quantum thermodynamic uncertainty relation for feedback control under jump measurement,
which provides the lower bound to the scaled variance of the number of jumps.
We find that the presence of feedback control can increase the accuracy of continuous measured systems, which is verified with numerical simulations. 
Moreover, we derive a quantum thermodynamic uncertainty relation for feedback control under homodyne detection. 
\end{abstract}
\maketitle

Cost, speed, and quality are representative trade-off elements in the world. It costs more to execute tasks faster and more accurately. The trade-offs between these elements are nowadays described quantitatively in the form of inequalities in thermodynamics and quantum mechanics.
The concept of a trade-off between cost and speed comes from the quantum speed limit (QSL) \cite{Deffner:2017:QSLReview}, first identified in 1945 \cite{Mandelstam:1945:QSL}.
This principle sets the absolute minimum amount of time necessary for a closed quantum system to transition from its initial state to its final one.
The QSL is not only being extended to different dynamics, but its applications in quantum computing \cite{Lloyd:2000:CompLimit}, quantum communication \cite{Bekenstein:1981:InfoTransfer,Murphy:2010:QSLchain}, and quantum thermodynamics \cite{Deffner:2010:GenClausius} are also being developed.
On the other hand, the trade-off between cost and quality was provided by the thermodynamic uncertainty relation (TUR), a concept founded in 2015 \cite{Barato:2015:UncRel} within the realm of classical stochastic thermodynamics \cite{Gingrich:2016:TUP,Horowitz:2019:TURReview}.
The TUR states that for a system to function with greater accuracy, it must incur higher thermodynamic costs, typically manifested as increased entropy production or dynamical activity.

Recently, TUR in quantum systems has been actively discussed beyond research in classical nonequilibrium systems \cite{Erker:2017:QClockTUR,Brandner:2018:Transport,Carollo:2019:QuantumLDP,Hasegawa:2020:QTURPRL,Guarnieri:2019:QTURPRR,Saryal:2019:TUR,Hasegawa:2020:TUROQS,Hasegawa:2021:QTURLEPRL,Vu:2021:QTURPRL,Monnai:2022:QTUR,Hasegawa:2023:BulkBoundaryBoundNC}.
In the TURs in quantum systems, 
it is particularly noteworthy to highlight how the quantum characteristics can be leveraged to enhance precision.
For instance, Ref.~\cite{Hasegawa:2020:QTURPRL} demonstrated how the precision can be amplified through the coherent dynamics of the Lindblad equation. Likewise, Ref.~\cite{Kalaee:2021:QTURPRE} confirmed that the accuracy can be increased by implementing quantum coherence.
These studies indicate that harnessing quantum properties can effectively enhance accuracy. In addition to exploiting these quantum properties, it appears feasible to boost precision through an externally manipulating system through quantum feedback \cite{Zhang:2017:QFeedbackReview}. In quantum feedback, the quantum operation for the next step is determined according to the output of the observation. Quantum feedback has been applied to quantum metrology \cite{Zhou:2018:HLECC}, quantum error correction \cite{Ahn:2003:ContQEC,Ahn:2004:FeedbackQEC}, and to name but a few.

Taking into account this background, in this Letter, we derive a quantum TUR for systems under continuous measurement feedback control \cite{Zhang:2017:QFeedbackReview}. 
Specifically, we employ the Markovian feedback control pioneered by Refs.~\cite{Wiseman:1993:Feedback,Wiseman:1994:Feedback}. 
By encoding the system dynamics and jump information into the matrix product state (MPS), we apply the quantum Cram\'er--Rao inequality to the feedback system.
We derive a quantum TUR whose upper bound comprises the quantum dynamical activity \cite{Hasegawa:2020:QTURPRL,Hasegawa:2023:BulkBoundaryBoundNC}. 
We show that, in the presence of feedback control, the accuracy of the continuously measured system can be increased. 
The quantum Cram{\'e}r-Rao inequality has the characteristic of being valid for any observable. Taking advantage of this property, we derive a quantum TUR valid for homodyne measurements.

\begin{figure}[t]
\includegraphics[width=8.5cm]{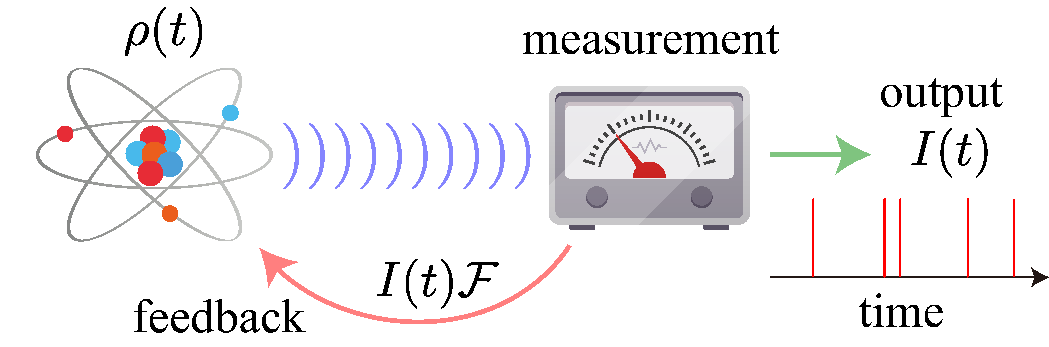} 
\caption{ 
Illustration of feedback control in continuous measurement. 
The measurement output $z(t)$ is used to control the system via $I(t)\mathcal{F}$. 
}
\label{fig:stochastic_process}
\end{figure}

\textit{Methods.---}In this Letter, we consider the continuous measurement formalism, which repeatedly monitors a quantum system's state (see Refs.~\cite{Annby:2022:PhDThesis,Landi:2023:CurFlucReview} for reviews).
Let
$\rho(t)$ be a density operator and
$\mathcal{H}$ be a superoperator defined by $\mathcal{H}\rho \equiv -i[H,\rho]$, which induces a unitary time-evolution with the Hamiltonian $H$. 
Let us begin with the following Lindblad equation \cite{Gorini:1976:GKSEquation,Lindblad:1976:Generators}:
\begin{align}
    \frac{d\rho}{dt}=\mathcal{L}\rho=\mathcal{H}\rho+\sum_{z=1}^{N_{C}}\mathcal{D}\left[L_{z}\right]\rho,
    \label{eq:GKSL_eq_def}
\end{align}
where
$\mathcal{L}$ is a Lindblad superoperator, 
$\mathcal{D}[L] \rho=L \rho L^{\dagger}-\frac{1}{2}\left\{L^{\dagger} L, \rho\right\}$ is the dissipator, and $N_C$ is the number of channels. 
Let us consider an infinitesimal time evolution of Eq.~\eqref{eq:GKSL_eq_def}. 
Upto the first order in $dt$, 
time evolution of Eq.~\eqref{eq:GKSL_eq_def} from $\rho(t)$ to $\rho(t+dt)$ can be represented by
\begin{align}
    \rho(t+dt)=e^{\mathcal{H}dt}\sum_{z=0}^{N_{C}}M_{z}\rho(t)M_{z}^{\dagger},
    \label{eq:Kraus_UM_def}
\end{align}
where $M_z$ are the Kraus operators defined by $M_{0}=1-\frac{1}{2}\sum_{z=1}^{N_{C}}L_{z}^{\dagger}L_{z}dt$ and $M_{z}=\sqrt{dt}L_{z}$ ($1 \le z \le N_C$). Note that when $z$ is continuous, the summation in Eq.~\eqref{eq:Kraus_UM_def} should be replaced by integration. 
Here $M_z$ ($1 \le z \le N_C$) corresponds to the jump induced by $L_z$ and $M_0$ to no-jump event. 
Equation~\eqref{eq:Kraus_UM_def} shows that the time evolution comprises two dynamics: a discontinuous jump induced by measurement $M_z$ and a unitary evolution induced by the Hamiltonian $H$. 
The Kraus representation of Eq.~\eqref{eq:Kraus_UM_def} shows a stochastic time evolution dependent on the measurement record $z$. 
Let $\rho_c(t)$ be a density operator conditioned on the measurement records $z$. At each time step within $[t,t+dt]$, the $z$th jump event is selected with probability $p_z(t) \equiv \mathrm{Tr}[M_z\rho_c(t)M_z^\dagger]$, which is followed by an update in the density operator $\rho_c$. The time evolution of $\rho_c(t)$ is described by
\begin{align}
    \rho_{c}(t+dt)=e^{\mathcal{H}dt}\frac{M_{z}\rho_{c}(t)M_{z}^{\dagger}}{p_{z}(t)}.
    \label{eq:rho_c_def}
\end{align}
This process is known as the unraveling of the quantum master equation, where the 
resulting time evolution of $\rho_c(t)$ is referred to as a quantum trajectory.
Equation~\eqref{eq:rho_c_def} shows that unraveled dynamics comprises alternate dynamics between continuous Hamiltonian dynamics $H$ and discontinuous jumps $L_z$. 
Let us introduce random variables $dN_z$, which are $1$ when a jump corresponding to the output $z$ occurs and $0$ otherwise. 
Its probability is $P(dN_{z}=1|\rho_{c}(t))=\mathrm{Tr}[\rho_{c}(t)L_{z}^{\dagger}L_{z}]$. 
Using $dN_z$, 
we can define the output current $I(t)=\sum_{z=1}^{N_C}\nu_{z}\frac{dN_{z}}{dt}$,
where $\nu_z$ is a real parameter representing the weight of each channel. 
Suppose that we are considering the dynamics ranges within $[0,\tau]$. 
Integrating $I(t)$ from $t=0$ to $t=\tau$, we obtain
\begin{align}
    N(\tau)=\int_{0}^{\tau}dt\;I(t)=\sum_{z}\nu_{z}N_{z}(\tau),
    \label{eq:N_tau_def}
\end{align}
where $N_z(\tau)$ quantifies the number of $z$th jumps within $[0,\tau]$. 
$N(\tau)$ in Eq.~\eqref{eq:N_tau_def} is the quantity of interest in a quantum TUR considered in the present Letter.

Next, we introduce the feedback in continuous measurement \cite{Wiseman:1993:Feedback,Wiseman:1994:Feedback}. 
Suppose that the output current $I(t)$ is fed back into the dynamics via
\begin{align}
    \rho_{c}(t+dt)=e^{\mathcal{H}dt}e^{I(t)\mathcal{F}dt}\frac{M_{z}\rho_{c}(t)M_{z}^{\dagger}}{p_{z}(t)},
    \label{eq:rho_c_FB_def}
\end{align}
where $\mathcal{F}$ is a superoperator representing the effect of the feedback to the dynamics.
$\mathcal{F}$ is defined via
$\mathcal{F}\rho \equiv -i[F,\rho]$,
where $F$ is an Hermitian operator. 
Note that the feedback is applied \textit{after} the measurement. 
This type of feedback is known as Markovian feedback, since no time delay is accompanied with the feedback. 
Upon averaging all possible quantum trajectories, Eq.~\eqref{eq:rho_c_FB_def} reduces to the known differential equation with respect to $\rho(t)$ \cite{Wiseman:1994:Feedback,Landi:2023:CurFlucReview}:
\begin{align}
    \frac{d\rho}{dt}=\mathcal{H}\rho+\sum_{z}\Bigl[e^{\nu_{z}\mathcal{F}}L_{z}\rho L_{z}^{\dagger}-\frac{1}{2}L_{z}^{\dagger}L_{z}\rho-\frac{1}{2}\rho L_{z}^{\dagger}L_{z}\Bigr],
    \label{eq:feedback_Lindblad_def}
\end{align}
where the superoperator $e^{\nu_z \mathcal{F}}$ is applied only after the jump, corresponding to $L_z$, occurs. 
Apparently $\nu_z=0$, Eq.~\eqref{eq:feedback_Lindblad_def} is identical to Eq.~\eqref{eq:GKSL_eq_def}. 
Equation~\eqref{eq:feedback_Lindblad_def} has been used in several problems, including stabilizing entanglement \cite{Carvalho:2007:FeedbackEnt}, quantum error correcting code \cite{Ahn:2003:ContQEC,Ahn:2004:FeedbackQEC}, and charging quantum battery \cite{Yao:2021:FeedbackQB}.

Matrix product state (MPS) is a mathematical model commonly used to represent many-body quantum states. The MPS is a type of tensor network state, which has been expanded to handle one-dimensional systems that exist in continuous state spaces \cite{Verstraete:2010:cMPS,Osborne:2010:Holography}. 
Using MPS, it becomes possible to gain a clear understanding of the continuous measurement. 
MPS has been applied to the study of continuous measurement in stochastic and quantum thermodynamics \cite{Garrahan:2010:QJ,Lesanovsky:2013:PhaseTrans,Garrahan:2016:cMPS}.
Specifically, the MPS representation was employed in TUR and QSL under continuous measurement \cite{Hasegawa:2020:QTURPRL,Hasegawa:2020:TUROQS,Hasegawa:2021:QTURLEPRL,Hasegawa:2023:BulkBoundaryBoundNC,Hasegawa:2022:FPTTURPRE}.

Let us derive a quantum TUR for systems under feedback control, described by Eq.~\eqref{eq:rho_c_FB_def}.
Therefore, in this Letter, we extend MPS representation to cases where there is feedback control. 
Let $K$ be a sufficiently large natural number and divide the time interval $[0,\tau]$ into $K$ equipartitioned intervals, where
$t_0 = 0$ and $t_{K} = \tau$. 
The time increment becomes $dt = \tau / K$. 
The Kraus representation for the system under feedback is
\begin{align}
    \rho(t+dt)=\sum_{z}U_{z}M_{z}\rho(t)M_{z}^{\dagger}U_{z}^{\dagger},
    \label{eq:Kraus_FB}
\end{align}
where $U_{z}\equiv e^{-iHdt}e^{-i\nu_{z}F}$ is a unitary operator corresponding to the Hamiltonian including feedback $I(t)F$. 
Equation~\eqref{eq:Kraus_FB} shows that the unitary feedback $U_z$ is applied after measuring $z$ corresponding to the operator $M_z$. 
Then the state at $t=\tau$ can be represented by applying Eq.~\eqref{eq:Kraus_FB} $K$ times:
\begin{align}
    \rho(\tau)&=\sum_{\boldsymbol{z}}U_{z_{K-1}}M_{z_{K-1}}\cdots U_{z_{1}}M_{z_{1}}U_{z_{0}}M_{z_{0}}\rho(0)\nonumber\\&\times M_{z_{0}}^{\dagger}U_{z_{0}}^{\dagger}M_{z_{1}}^{\dagger}U_{z_{1}}^{\dagger}\cdots M_{z_{K-1}}^{\dagger}U_{z_{K-1}}^{\dagger},
    \label{eq:rho_FB_tau}
\end{align}
where $\boldsymbol{z} \equiv [z_0,\ldots,z_{K-1}]$. 
From Eq.~\eqref{eq:rho_FB_tau},  we can define the corresponding MPS state as follows:
\begin{align}
    \ket{\Psi(\tau)}\equiv\sum_{\boldsymbol{z}}U_{z_{K-1}}M_{z_{K-1}}\cdots U_{z_{1}}M_{z_{1}}U_{z_{0}}M_{z_{0}}\ket{\psi_{S}(0)}\otimes\ket{\boldsymbol{z}}.
    \label{eq:MPS_FB_def}
\end{align}
$\ket{\Psi(\tau)}$ encodes information of jump events into the field $\ket{\boldsymbol{z}}$.
The MPS in Eq.~\eqref{eq:MPS_FB_def} plays a central role in this Letter, as all the information of the continuous measurement can be encoded in a pure state $\ket{\Psi(\tau)}$. 

Next, we consider a parameter inference in the continuous measurement under feedback control. 
Quantum information holds a crucial position in shaping the uncertainty relations inherent in quantum systems. 
Let $\theta$ be a parameter of interest ($\theta \in \mathbb{R}$). 
Suppose that $H$, $L_z$, and $F$ are parametrized as
$H(\theta)$, $L_z(\theta)$, and $F(\theta)$, respectively. 
Without loss of generality, we assume that $H(\theta=0)=H$, $L_z(\theta=0)=L_z$, and $F(\theta=0)=F$. 
Let us consider inferring $\theta$ from the measurement of the output $I(t)$,
which is generated by the parametrized model [$H(\theta)$, $L_z(\theta)$, and $F(\theta)$]. 
The quantum Fisher information for the conventional continuous measurement was studied in Ref.~\cite{Gammelmark:2014:QCRB}. 
Here, we extend this quantum Fisher information
calculation \cite{Gammelmark:2014:QCRB} to the system in the presence of quantum feedback. 
Let us define $M_{0}(\theta)=1-\frac{1}{2}\sum_{z=1}^{N_{C}}L_{z}^{\dagger}(\theta)L_{z}(\theta)dt$ and $M_{z}(\theta)=\sqrt{dt}L_{z}(\theta)$ ($1 \le z \le N_C$)
and $U_{z}(\theta)\equiv e^{-iH(\theta)dt}e^{-i\nu_{z}F(\theta)dt}$. 
Moreover, let $\ket{\Psi(\tau;\theta)}$ be MPS [Eq.~\eqref{eq:MPS_FB_def}], whose operators $M_z$ and $U_z$ are replaced with $M_z(\theta)$ and $U_z(\theta)$, respectively. 
Since $\ket{\Psi(\tau;\theta)}$ in Eq.~\eqref{eq:MPS_FB_def} is a pure state in the composite space comprising the system and the environment (field),
we can represent the quantum Fisher information as 
\begin{align}
    \mathcal{I}(\tau;\theta)=\frac{8}{d\theta^{2}}(1-|\braket{\Psi(\tau;\theta+d\theta)|\Psi(\tau;\theta)}|).
    \label{eq:QFI_def}
\end{align}
Consider the quantum Cram\'er-Rao inequality. 
Let $\Theta$ be an observable of the continuous measurement. Then the quantum Cram\'er--Rao inequality holds:
\begin{align}
    \frac{\mathrm{Var}_{\theta}[\Theta]}{\left(\partial_{\theta}\braket{\Theta}_{\theta}\right)^{2}}\ge\frac{1}{\mathcal{I}(\tau;\theta)}.
    \label{eq:QCR_bound}
\end{align}
Recall that $\left|\braket{\Psi(\tau;\phi)|\Psi(\tau;\theta)}\right|=\mathrm{Tr}_{SF}\left[\ket{\Psi(\tau;\theta)}\bra{\Psi(\tau;\phi)}\right]=\mathrm{Tr}_{F}\left[\varrho(\tau;\theta,\phi)\right]$, where $\varrho(t;\theta,\phi)\equiv\mathrm{Tr}_{F}\left[\ket{\Psi(\tau;\theta)}\bra{\Psi(\tau;\phi)}\right]$. 
From the MPS representation of Eq.~\eqref{eq:MPS_FB_def},
it can be shown that $\varrho(t;\theta,\phi)$ obeys the two-sided Lindblad equation \cite{Gammelmark:2014:QCRB}. 
For simplicity and ease of notation, we may denote $\varrho(t;\theta,\phi)$ as $\varrho(t)$.
The following relation holds for the feedback system:
\begin{align}
    \varrho(t+dt)&=\sum_{z}U_{z}(\theta)M_{z}(\theta)\varrho(t)M_{z}^{\dagger}(\phi)U_{z}^{\dagger}(\phi)\label{eq:two_sided_Kraus1}
\end{align}
Let us define the \textit{two-sided} superoperators $\mathcal{H}(\theta,\phi)\rho\equiv-i[H(\theta)\rho-\rho H(\phi)]$ and $\mathcal{F}(\theta,\phi)\rho\equiv-i[F(\theta)\rho-\rho F(\phi)]$,
which are two-sided variants of $\mathcal{H}$ and $\mathcal{F}$ defined above. 
Solving Eq.~\eqref{eq:two_sided_Kraus1}, we derive the following differential equation:
\begin{align}
    \frac{d\varrho}{dt}&=\mathcal{H}(\theta,\phi)\varrho+\sum_{z}\Bigl[e^{\nu_{z}\mathcal{F}(\theta,\phi)}L_{z}(\theta)\varrho L_{z}^{\dagger}(\phi)\nonumber\\&-\frac{1}{2}L_{z}^{\dagger}(\theta)L_{z}(\theta)\varrho-\frac{1}{2}\varrho L_{z}^{\dagger}(\phi)L_{z}(\phi)\Bigr].
    \label{eq:twosided_feedback_Lindblad_eq}
\end{align}
When $\theta = \phi$, Eq.~\eqref{eq:twosided_feedback_Lindblad_eq} reduces to the feedback Lindblad equation of Eq.~\eqref{eq:feedback_Lindblad_def}. 
Calculating Eq.~\eqref{eq:twosided_feedback_Lindblad_eq} from the initial density operator $\varrho(0) = \ket{\psi(0)}\bra{\psi(0)}$ with 
$\phi = \theta + d\theta$ yields the quantum Fisher information via Eq.~\eqref{eq:QFI_def}. 

We next derive a quantum TUR under feedback using the quantum Cram\'er-Rao inequality following Ref.~\cite{Hasegawa:2020:QTURPRL}. 
Suppose the following parameterization:
\begin{align}
    H(\theta)= (1+\theta)H,
    L_z(\theta)=\sqrt{1+\theta}L_z,
    F(\theta) = F.
    \label{eq:HLF_scaling}
\end{align}
With the scaling of Eq.~\eqref{eq:HLF_scaling}, the Lindblad equation is given by Eq.~\eqref{eq:twosided_feedback_Lindblad_eq} with $\phi = \theta$.
This parametrized Lindblad equation is exactly the same as the original equation Eq.~\eqref{eq:feedback_Lindblad_def} except for the time scale. 
Therefore, $\braket{N(\tau)}_\theta = (1+\theta)\braket{N(\tau)}$,
where $\braket{\bullet}_\theta$ denotes the expectation calculated with the parameterization of Eq.~\eqref{eq:HLF_scaling}, and 
$\braket{\bullet} = \braket{\bullet}_{\theta=0}$ is the expectation of the original (unparameterized) Lindblad equation. 
Then we derive a quantum TUR under feedback control from Eq.~\eqref{eq:QCR_bound}:
\begin{align}
    \frac{\mathrm{Var}[N(\tau)]}{\braket{N(\tau)}^{2}}\ge\frac{1}{\mathcal{B}_{\mathrm{jmp}}^{\mathrm{fb}}(\tau)},
    \label{eq:QTUR_fb}
\end{align}
where $\mathcal{B}_{\mathrm{jmp}}^{\mathrm{fb}}(\tau)$ is the quantum dynamical activity under feedback control 
$\mathcal{B}_{\mathrm{jmp}}^{\mathrm{fb}}(\tau) = \mathcal{I}(\tau;\theta = 0)$, where the quantum Fisher information is calculated with the parametrization of Eq.~\eqref{eq:HLF_scaling}.  
Equation~\eqref{eq:QTUR_fb} is the main result of this Letter. 
Reference~\cite{Hasegawa:2020:QTURPRL} evaluated the quantum dynamical activity without feedback for $\tau \to \infty$ after Ref.~\cite{Gammelmark:2014:QCRB}.
Similarly, we can evaluate $\mathcal{B}_{\mathrm{jmp}}^{\mathrm{fb}}(\tau)$ for $\tau \to \infty$ using the Choi-Jamio{\l}kowski isomorphism \cite{Supp:2023:FeedbackQTUR}.

We have considered feedback control in the jump measurement. 
Since the Lindblad equation is invariant under the gauge transformation,
we can consider different continuous measurements other than the jump measurement \cite{Landi:2023:CurFlucReview}. 
We can derive homodyne detection using a method that involves the continuous application of weak Gaussian measurements \cite{Jacobs:2006:ContMeas,Landi:2023:CurFlucReview}.
Following Refs.~\cite{Jacobs:2006:ContMeas,Landi:2023:CurFlucReview},
for homodyne detection, the measurement operator is defined by
\begin{align}
    M_z=\left(\frac{2 \lambda d t}{\pi}\right)^{\frac{1}{4}} e^{-\lambda d t(z-Y)^2},
    \label{eq:Mz_homodyne_def}
\end{align}
where $Y$ is an Hermitian operator and 
$\lambda$ denotes the strength of the measurement.
Then, 
the measurement output $z$ can be approximated by $z=\mathrm{Tr}[\rho_c(t)Y]+\frac{dW}{2\sqrt{\lambda}dt}$,
where $dW$ is the Wiener increment satisfying $\braket{dW}=0$
and $\braket{dW^{2}}=dt$. 
For $\lambda \to \infty$, the measurement reduces to the projective measurement, while $\lambda \to 0$ corresponds to weak measurements which only slightly disturb the system. 
Taking the avarage with respect to the measurement records $z$, 
the Wiseman-Milburn equation can be reproduced \cite{Wiseman:1993:Feedback}:
\begin{align}
    \frac{d\rho}{dt}=\mathcal{H}\rho+\lambda\mathcal{D}[Y]\rho+\frac{1}{2}\mathcal{F}\{Y,\rho\}+\frac{1}{8\lambda}\mathcal{F}^{2}\rho.
    \label{eq:Wiseman_Milburn_def}
\end{align}
In Eq.~\eqref{eq:Wiseman_Milburn_def}, the first and second terms are identical to those in the Lindblad equation. The third and fourth terms give insight into the ways in which the system is affected by the feedback. 
We next show that we can consider feedback control in homodyne detection as well. 
As in the jump measurement, we need to calculate the two-sided variant of the Wiseman-Milburn equation to calculate the quantum dynamical activity. 
The two-sided Wiseman-Milburn equation is given by (see Ref.~\cite{Supp:2023:FeedbackQTUR} for details)
\begin{align}
    \frac{d\varrho}{dt}&=\frac{1}{8\lambda}\mathcal{F}(\theta,\phi)^{2}\varrho+\frac{1}{2}\mathcal{F}(\theta,\phi)\left(\varrho Y(\phi)+Y(\theta)\varrho\right)\nonumber\\&+\mathcal{H}(\theta,\phi)\varrho-\frac{1}{2}\lambda\varrho Y(\phi)^{2}-\frac{1}{2}\lambda Y(\theta)^{2}\varrho\nonumber\\&+\lambda Y(\theta)\varrho Y(\phi).
    \label{eq:TS_WisemanMilburn_eq}
\end{align}
In the right-hand side of Eq.~\eqref{eq:TS_WisemanMilburn_eq}, the first and the second terms signify the effects of feedback control. 
Without these two terms, 
Eq.~\eqref{eq:TS_WisemanMilburn_eq} reduces to the two-sided Lindblad equation in Ref.~\cite{Gammelmark:2014:QCRB}. 
Apparently, for $\theta=\phi=0$, Eq.~\eqref{eq:TS_WisemanMilburn_eq} is identical to the Wiseman-Milburn equation [Eq.~\eqref{eq:Wiseman_Milburn_def}]. 
Now we are interested in the time-integrated output $Z(\tau)\equiv\int_{0}^{\tau}z(t)dt$. 
Suppose the following parameterization for the homodyne detection:
\begin{align}
    H(\theta)= (1+\theta)H,
    L_z(\theta)=\sqrt{1+\theta}L_z,
    F(\theta) =\sqrt{1+\theta} F.
    \label{eq:HLF_scaling_hom}
\end{align}
The dynamics under the parametrization of Eq.~\eqref{eq:HLF_scaling_hom} is given by Eq.~\eqref{eq:TS_WisemanMilburn_eq} with $\phi = \theta$. 
Under the parameterization of Eq.~\eqref{eq:HLF_scaling_hom}, the average of $Z(\tau)$ scales as
$\braket{Z(\tau)}_{\theta}=\int_{0}^{\tau}\braket{\sqrt{1+\theta}Y}dt=\sqrt{1+\theta}\braket{Z(\tau)}$,
where $\braket{\bullet}=\braket{\bullet}_{\theta=0}$ is the expectation with respect to the original dynamics.
Therefore, from the quantum Cram{\'e}r-Rao inequality, we obtain
\begin{align}
    \frac{\mathrm{Var}[Z(\tau)]}{\braket{Z(\tau)}^{2}}\ge\frac{1}{4\mathcal{B}_{\mathrm{hom}}^{\mathrm{fb}}(\tau)},
    \label{eq:QTUR_fb_homodyne}
\end{align}
where 
$\mathcal{B}_{\mathrm{hom}}^{\mathrm{fb}}(\tau) = \mathcal{I}(\tau;\theta = 0)$, where the quantum Fisher information is calculated with
the two-sided Lindblad equation [Eq.~\eqref{eq:TS_WisemanMilburn_eq}] with the parametrization of Eq.~\eqref{eq:HLF_scaling_hom}.  
Without feedback, on replacing $L = \sqrt{\lambda}Y$,
$\mathcal{B}^\mathrm{fb}_\mathrm{jmp}(\tau)$ in Eq.~\eqref{eq:QTUR_fb} and $\mathcal{B}^\mathrm{fb}_\mathrm{hom}(\tau)$ in Eq.~\eqref{eq:QTUR_fb_homodyne} become identical,
because the two-sided Lindblad equations for these two cases agree.

\begin{figure}
\includegraphics[width=8cm]{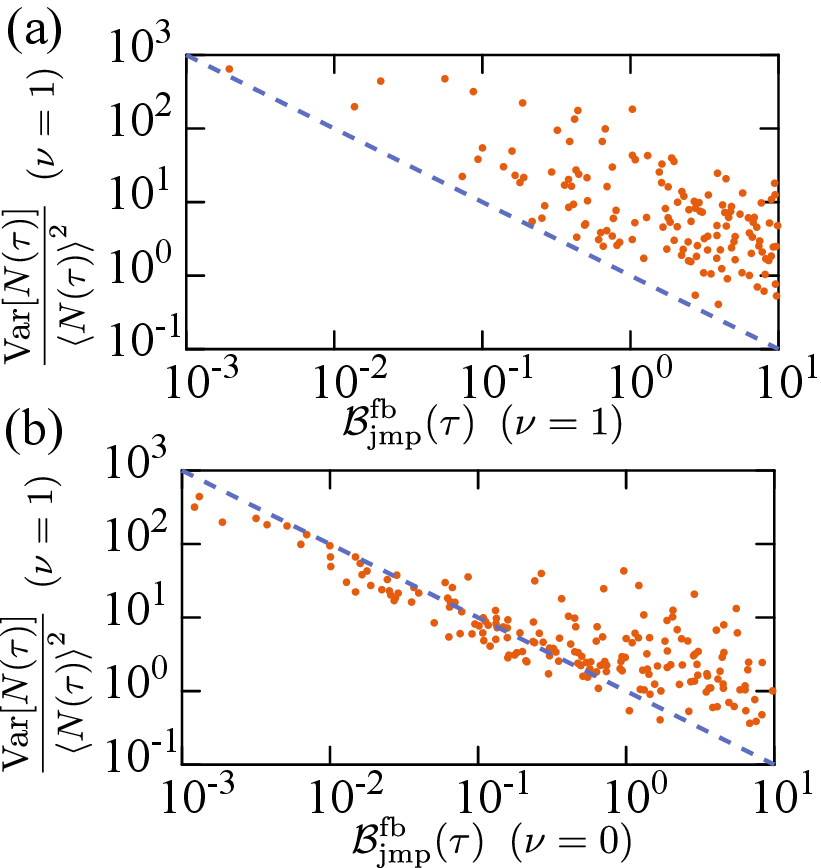} 
\caption{
Numerical simulation of the quantum TUR under feedback in jump measurement.
(a) and (b) 
Precision $\mathrm{Var}[N(\tau)]/\braket{N(\tau)}^{2}$ as a function of $\mathcal{B}^\mathrm{fb}_\mathrm{jmp}(\tau)$ for random realizations, where
random realizations are plotted by circles and the solid line denotes $1 / \mathcal{B}^\mathrm{fb}_\mathrm{jmp}(\tau)$. 
In (a), the precision obtained with $\nu=1$ is plotted as a function $\mathcal{B}^\mathrm{fb}_\mathrm{jmp}(\tau)$ with $\nu = 1$, which should satisfy Eq.~\eqref{eq:QTUR_fb}, i.e., all points should be located above the solid line. 
On the other hand, in (b), the precision with $\nu=1$ (with feedback) is plotted as a function $B$ with $\nu = 0$ (without feedback), which is \textit{not} expected to satisfy Eq.~\eqref{eq:QTUR_fb}. 
In (a) and (b), the parameter ranges for the random realizations are 
$\Delta \in [0.1,3.0]$, $\Omega \in [0.1,3.0]$, $\kappa \in [0.1,3.0]$, and $\tau \in [0.1,3.0]$. 
}
\label{fig:bound_plot}
\end{figure}

\textit{Numerical simulation.---}We perform numerical simulations to validate the quantum TUR under feedback control [Eq.~\eqref{eq:QTUR_fb}]. 
We consider a two-level atom driven by a classical laser field.
Let $\ket{e}$ and $\ket{g}$ be excited and ground states, respectively.
whose Hamiltonian and jump operator are given by
\begin{align}
    H=\Delta\ket{e}\bra{e}+\frac{\Omega}{2}\left(\ket{e}\bra{g}+\ket{g}\bra{e}\right),
    \label{eq:Rabi_Hamiltonian}
\end{align}
and $L = \sqrt{\kappa}\ket{g}\bra{e}$. 
Here, $\Delta$, $\Omega$, and $\kappa$ are model parameters. 
We are interested in fluctuations in the number of jump events $N(\tau)$
within the interval $[0,\tau]$. 
We randomly determine the model parameters and the time duration $\tau$ (see the caption of Fig.~\ref{fig:bound_plot} for details).
Then, we calculate $\mathrm{Var}[N(\tau)] / \braket{N(\tau)}^2$ with the simulation, and evaluate the quantum dynamical activity $\mathcal{B}^\mathrm{fb}_\mathrm{hom}(\tau)$. 
For the feedback operator, we employ
\begin{align}
    F = \ket{e}\bra{g}+\ket{g}\bra{e}.
    \label{eq:F_for_Rabi}
\end{align}
The strength of the feedback is control by $\nu$ defined above. 

Figure~\ref{fig:bound_plot} shows results of the numerical simulation for $\nu = 1$, where points denote $\mathrm{Var}[N(\tau)]/\braket{N(\tau}^{2}$ as a function of $\mathcal{B}^\mathrm{fb}_\mathrm{jmp}(\tau)$ for the random realizations
and the dashed line denotes the lower bound of Eq.~\eqref{eq:QTUR_fb}. 
Since all the points are located above the dashed line, we numerically confirm that the quantum TUR given by Eq.~\eqref{eq:QTUR_fb} holds for the feedback quantum dynamics.
Next, we see whether precision $\mathrm{Var}[N(\tau)]/\braket{N(\tau}^{2}$ can be improved in the presence of feedback $\nu \mathcal{F}$. 
In Fig.~\ref{fig:bound_plot}(b), we plot $\mathrm{Var}[N(\tau)]/\braket{N(\tau}^{2}$ for $\nu = 1$ as a function of $\mathcal{B}^\mathrm{fb}_\mathrm{jmp}(\tau)$ for $\nu = 0$, which is the quantum dynamical activity \textit{without} feedback. 
As in Fig.~\ref{fig:bound_plot}(a), the points denote the random realizations and the solid line is $1/\mathcal{B}^\mathrm{fb}_\mathrm{jmp}(\tau)$, where $\mathcal{B}^\mathrm{fb}_\mathrm{jmp}(\tau)$ calculated with $\nu = 0$.
Some points are below the dashed line, which implies that the precision is improved in the presence of feedback.

\textit{Conclusion.---}This Letter presents the quantum TUR for the continuous measurement under feedback control. We considered jump and diffusion measurements and derived the quantum dynamical activity for the two cases. We showed that the presence of feedback control increases the quantum dynamical activity, which in turn improves the precision of counting observable in the continuous measurement. 
Feedback is a central technique in quantum engineering.
Therefore, it is expected that this research will clarify accurate design guidelines for quantum thermodynamic systems.

\begin{acknowledgments}
This work was supported by JSPS KAKENHI Grant Number JP22H03659.
\end{acknowledgments}

\end{document}


\title{Supplementary Material for\\ ``Quantum Thermodynamic Uncertainty Relation under Feedback Control''}
\author{Yoshihiko Hasegawa}
\email{hasegawa@biom.t.u-tokyo.ac.jp}
\affiliation{Department of Information and Communication Engineering, Graduate
School of Information Science and Technology, The University of Tokyo,
Tokyo 113-8656, Japan}

\maketitle
This supplementary material describes the calculations introduced in the main text. The numbers of the equations and the figures are prefixed with S (e.g., Eq.~(S1) or Fig.~S1). Numbers without this prefix (e.g., Eq.~(1) or Fig.~1) refer to items in the main text.

\section{Two-sided equation for jump measurement\label{sec:TS_Jump_equation}}

In this section, we show a detailed derivation of the two-sided equation for jump measurement. 
Equation~\twoUsidedUKrausI{} for jump measurement is calculated as follows:
\begin{align}
    \varrho(t+dt)&=\sum_{z}U_{z}(\theta)M_{z}(\theta)\varrho(t)M_{z}^{\dagger}(\phi)U_{z}^{\dagger}(\phi)\nonumber\\&=e^{\mathcal{H}(\theta,\phi)dt}\sum_{z}e^{\nu_{z}\mathcal{F}(\theta,\phi)}M_{z}(\theta)\varrho(t)M_{z}^{\dagger}(\phi),
    \label{eq:two_sided_Kraus2}
\end{align}
where the superoperators $\mathcal{H}(\theta,\phi)$ and $\mathcal{F}(\theta,\phi)$ are defined in the main text by
\begin{align}
\mathcal{H}(\theta,\phi)\rho=-i[H(\theta)\rho-\rho H(\phi)],\label{eq:mathcal_H_twosided}\\
    \mathcal{F}(\theta,\phi)\rho=-i[F(\theta)\rho-\rho F(\phi)].
    \label{eq:mathcal_F_twosided}
\end{align}
The term in Eq.~\eqref{eq:two_sided_Kraus2} is evaluated as follows:
\begin{align}
    \sum_{z}e^{\nu_{z}\mathcal{F}(\theta,\phi)}M_{z}(\theta)\varrho(t)M_{z}^{\dagger}(\phi)&=M_{0}(\theta)\varrho(t)M_{0}(\phi)+\sum_{z=1}^{N_{C}}e^{\nu_{z}\mathcal{F}(\theta,\phi)}M_{z}(\theta)\varrho(t)M_{z}(\phi)\nonumber\\&=\varrho(t)+\sum_{z=1}^{N_{C}}\left[e^{\nu_{z}\mathcal{F}(\theta,\phi)}L_{z}(\theta)\varrho(t)L_{z}(\phi)dt-\frac{1}{2}\varrho(t)L_{z}^{\dagger}(\phi)L_{z}(\phi)dt-\frac{1}{2}L_{z}^{\dagger}(\theta)L_{z}(\theta)\varrho(t)dt\right].
    \label{eq:two_sided_Kraus3}
\end{align}
Applying $e^{\mathcal{H}(\theta,\phi)dt}$ to Eq.~\eqref{eq:two_sided_Kraus3}, we obtain 
\begin{align}
    \frac{d\varrho}{dt}=\mathcal{H}(\theta,\phi)\varrho+\sum_{z}\left[e^{\nu_{z}\mathcal{F}(a,b)}L_{z}(\theta)\varrho(t)L_{z}^{\dagger}(\phi)-\frac{1}{2}L_{z}^{\dagger}(\theta)L_{z}(\theta)\varrho(t)-\frac{1}{2}\varrho(t)L_{z}^{\dagger}(\phi)L_{z}(\phi)\right],
    \label{eq:jump_twosided_def}
\end{align}
which is Eq.~\twosidedUfeedbackULindbladUeq{}.
Recall that the parametrized operators $H(\theta)$, $L_z(\theta)$, and $F(\theta)$ reduce to $H$, $L_z$, and $F$, respectively, for $\theta=0$. 
Taking $\theta = \phi = 0$, Eq.~\eqref{eq:jump_twosided_def} reduces to the well-known equation:
\begin{align}
    \frac{d\varrho}{dt}=\mathcal{H}\varrho+\sum_{z}\left[e^{\nu_{z}\mathcal{F}}L_{z}\varrho(t)L_{z}^{\dagger}-\frac{1}{2}L_{z}^{\dagger}L_{z}\varrho(t)-\frac{1}{2}\varrho(t)L_{z}^{\dagger}L_{z}\right],
    \label{eq:feedback_Lindblad}
\end{align}
which corresponds to Eq.~\feedbackULindbladUdef{} in the main text.

\section{Two-sided equation for homodyne detection\label{sec:TS_WM_equation}}

We provide a derivation of the two-sided Wiseman-Milburn equation. 
The homodyne detection admits the Gaussian positive operator-valued measure (POVM) representation \cite{Jacobs:2006:ContMeas}.
We employ this representation to derive the two-sided Wiseman-Milburn equation. 
Let us define the Kraus operator parameterized by $\theta$:
\begin{align}
    M_z(\theta)=\left(\frac{2 \lambda d t}{\pi}\right)^{\frac{1}{4}} e^{-\lambda d t(z-Y(\theta))^2}.
    \label{eq:par_Mz_homodyne_def}
\end{align}
We now calculate the following map:
\begin{align}
    \varrho(t+dt)&=e^{\mathcal{H}(\theta,\phi)dt}\int dz\,M_{z}(\theta)\varrho(t)M_{z}^{\dagger}(\phi)\nonumber\\&=e^{\mathcal{H}(\theta,\phi)dt}\int dz\,\mathfrak{p}_{z}(t)\frac{M_{z}(\theta)\varrho(t)M_{z}^{\dagger}(\phi)}{\mathfrak{p}_{z}(t)}\nonumber\\&=e^{\mathcal{H}(\theta,\phi)dt}\Braket{\frac{M_{z}(\theta)\varrho(t)M_{z}^{\dagger}(\phi)}{\mathfrak{p}_{z}(t)}},
    \label{eq:homodyne_CPTP}
\end{align}
where the expectation in the last line is with respect to the Wiener process. 
Recall that the reference probability $\mathfrak{p}_z(t)$ in Eq.~\eqref{eq:homodyne_CPTP} can be chosen arbitrarily. 
Let us define $\mathfrak{p}_z(t)$ as follows:
\begin{align}
    \mathfrak{p}_{z}(t)\equiv \mathrm{Tr}\left[M_{z}\sigma(t)M_{z}^{\dagger}\right],
    \label{eq:pz_def}
\end{align}
where $\sigma(t)$ is the operator defined by
\begin{align}
    \sigma(t)\equiv\frac{\varrho(t)+\varrho^{\dagger}(t)}{\mathrm{Tr}[\varrho(t)+\varrho^{\dagger}(t)]}.
    \label{eq:sigma_op_def}
\end{align}
Straight-forward calculation yields
\begin{align}
    \mathfrak{p}_{z}(t)=\left(\frac{2\lambda dt}{\pi}\right)^{\frac{1}{2}}\sum_{y}e^{-2\lambda dt(z-y)^{2}}\braket{y|\sigma|y}.
    \label{eq:pz_2}
\end{align}
Let us calculate the mean and variance of $z$:
\begin{align}
    \int dz\,z\mathfrak{p}_{z}(t)&=\sum_{y}y\braket{y|\sigma|y},\label{eq:z_mean}\\
    \mathrm{\mathrm{Var}}(z)&=\frac{1}{4\lambda dt}+\braket{Y^{2}}-\braket{Y}^{2}\simeq\frac{1}{4\lambda dt}.
    \label{eq:z_var}
\end{align}
From Eqs.~\eqref{eq:z_mean} and \eqref{eq:z_var}, we can approximate $z$ by
\begin{align}
    z=\mathrm{Tr}[Y\sigma(t)]+\frac{dW}{2\sqrt{\lambda}dt},
    \label{eq:z_Wiener}
\end{align}
where $dW$ is the Wiener process. Using Eq.~\eqref{eq:z_Wiener} and $dW^2 = dt$, we derive
\begin{align}
    \frac{M_{z}(\theta)\rho(t)M_{z}^{\dagger}(\phi)}{\mathfrak{p}_{z}(t)}&=\rho-2\rho\sqrt{\lambda}dW\langle Y\rangle_{\sigma}+\sqrt{\lambda}dW\rho Y(\phi)+\sqrt{\lambda}dWY(\theta)\rho\nonumber\\&-\frac{1}{2}\lambda dt\rho Y(\phi)^{2}-\frac{1}{2}\lambda dtY(\theta)^{2}\rho+\lambda Y(\theta)\rho Y(\phi)dt.
    \label{eq:Kz_pz}
\end{align}
Taking average in Eq.~\eqref{eq:Kz_pz}, we obtain
\begin{align}
    \int dz\,M_{z}(\theta)\varrho(t)M_{z}(\phi)=\left\langle \frac{M_{z}(\theta)\varrho(t)M_{z}^{\dagger}(\phi)}{\mathfrak{p}_{z}(t)}\right\rangle =\varrho-\frac{1}{2}\lambda dt\varrho Y(\phi)^{2}-\frac{1}{2}\lambda dtY(\theta)^{2}\varrho+\lambda Y(\theta)\varrho Y(\phi)dt.
    \label{eq:Kz_pz_average}
\end{align}
Applying $e^{\mathcal{H}(\theta,\phi)t}$ to Eq.~\eqref{eq:Kz_pz_average},
we have
\begin{align}
    \frac{d\varrho}{dt}=e^{\mathcal{H}(\theta,\phi)dt}\rho-\frac{1}{2}\lambda dt\varrho Y(\phi)^{2}-\frac{1}{2}\lambda dtY(\theta)^{2}\varrho+\lambda Y(\theta)\varrho Y(\phi)dt,
    \label{eq:Kz_pz_average2}
\end{align}
which is identical to the conventional two-sided Lindblad equation. 
In fact, let us take $\sqrt{\lambda}Y(\theta) = L(\theta)$, where $L(\theta)$ is the jump operator. Then we obtain
\begin{align}
    \frac{d\varrho}{dt}=e^{\mathcal{H}(\theta,\phi)dt}\rho-\frac{1}{2}dt\varrho L(\phi)^{2}-\frac{1}{2}dtL(\theta)^{2}\varrho+L(\theta)\varrho L(\phi)dt.
    \label{eq:Kz_pz_average3}
\end{align}

Next, we consider the feedback by adding $e^{\nu \mathcal{F}}$. 
From the Taylor expansion, we have
\begin{align}
e^{\mathcal{H}(\theta,\phi)dt}e^{z\mathcal{F}(\theta,\phi)}=1+\mathcal{H}(\theta,\phi)dt+\braket{Y}_{\sigma}\mathcal{F}(\theta,\phi)dt+\frac{dW}{\sqrt{4\lambda}}\mathcal{F}(\theta,\phi)+\frac{dt}{8\lambda}\mathcal{F}(\theta,\phi)^{2}.
    \label{eq:exp_H_zF}
\end{align}
In ths presence of the feedback, the quantum channel is given by
\begin{align}
    \varrho(t+dt)=\left<e^{\mathcal{H}(\theta,\phi)dt}e^{z\mathcal{F}(\theta,\phi)}\frac{M_{z}(\theta)\varrho(t)M_{z}^{\dagger}(\phi)}{\mathfrak{p}_{z}(t)}\right>.
    \label{eq:CPTP_feedback}
\end{align}
Specifically, evaluating the expectation with respect to the Wiener process, we obtain
\begin{align}
    \left\langle e^{\mathcal{H}(\theta,\phi)dt}e^{z\mathcal{F}(\theta,\phi)}\frac{M_{z}(\theta)\varrho(t)M_{z}^{\dagger}(\phi)}{\mathfrak{p}_{z}(t)}\right\rangle &=\rho+\mathcal{H}(\theta,\phi)\rho dt+\frac{dt}{8\lambda}\mathcal{F}(\theta,\phi)^{2}\rho+\frac{dt}{2}\mathcal{F}(\theta,\phi)\rho Y(\phi)+\frac{dt}{2}\mathcal{F}(\theta,\phi)Y(\theta)\rho\nonumber\\&-\frac{1}{2}\lambda dt\rho Y(\phi)^{2}-\frac{1}{2}\lambda dtY(\theta)^{2}\rho+\lambda Y(\theta)\rho Y(\phi)dt,
    \label{eq:two_sided_MW_equation}
\end{align}
which is Eq.~\TSUWisemanMilburnUeq{} in the main text.

\section{Asymptotic calculation of quantum dynamical activity\label{sec:QDA_calculation}}

In this section, we show the asymptotic expression of $\mathcal{B}_\mathrm{jmp}^\mathrm{fb}(\tau)$ for $\tau \to \infty$. 
Because the asymptotic expression employs the vectorization via Choi-Jamio{\l}kowski isomorphism, we first briefly introduce its concept. 
Consider the density operator $\rho$:
\begin{align}
    \rho = \sum_{i,j}\rho_{ij}\ket{i}\bra{j}.
    \label{eq:rho_def}
\end{align}
where $\ket{i}$ is an orthonormal basis. 
Using the vectorization, $\rho$ can be converted to
\begin{align}
    \dblket{\rho}\equiv\sum_{i,j}\rho_{ij}\ket{j}\otimes\ket{i},
    \label{eq:Liouville_rho_def}
\end{align}
which corresponds to the column stacking of the matrix. 
When $\rho$ in Eq.~\eqref{eq:rho_def} is an $n\times n$ matrix,
$\dblket{\rho}$ becomes an $n^2$-dimensional column vector. 
The bra vector of $\dblket{\rho}$ is defined by
\begin{align}
    \dblbra{\rho} \equiv \dblket{\rho}^\dagger.
    \label{eq:dblbra_def}
\end{align}
With this notation, the inner product becomes the Hilbert-Schmidt inner product:
\begin{align}
    \mathrm{Tr}\left[A^{\dagger}B\right]=\dblbraket{A\mid B}.
    \label{eq:inner_product_def}
\end{align}
The following relation is useful for calculation of the Lindblad equation:
\begin{align}
    \dblket{ABC}=(C^{\top}\otimes A)\dblket{B},
    \label{eq:ABC_dblket}
\end{align}
where $A$, $B$, and $C$ are arbitrary matrices. 
Using Eq.~\eqref{eq:ABC_dblket}, the Lindblad super-operator [Eq.~\GKSLUeqUdef{}] becomes
\begin{align}
    \frac{d}{dt}\dblket{\rho}=\mathcal{L}\dblket{\rho},
    \label{eq:GKSL_vec_def}
\end{align}
where $\mathcal{L}$ is vectorized as follows:
\begin{align}
    \mathcal{L}=-i\left(\mathbb{I}\otimes H-H^{\top}\otimes\mathbb{I}\right)+\sum_{z}\left[e^{\nu\mathcal{F}}L_{z}^{*}\otimes L_{z}-\frac{1}{2}\mathbb{I}\otimes L_{z}^{\dagger}L_{z}-\frac{1}{2}\left(L_{z}^{\dagger}L_{z}\right)^{\top}\otimes\mathbb{I}\right].
    \label{eq:L_sop_def}
\end{align}
Moreover, the steady-state density matrix of the Lindblad equation is the eigenvector of $\mathcal{L}$ corresponding to the vanishing eigenvalue: 
\begin{align}
    \mathcal{L}\dblket{\rho_{\mathrm{ss}}}=0.
    \label{eq:rhoss_L_vec}
\end{align}

Based on Ref.~\cite{Gammelmark:2014:QCRB}, we 
have calculated the quantum dynamical activity (without feedback)
for $\tau \to \infty$ \cite{Hasegawa:2020:QTURPRL}.
Following Ref.~\cite{Hasegawa:2020:QTURPRL}, we here show how we can compute the quantum dynamical activity in the presence of feedback control. 
Using the vectorization, we can represent the quantum dynamical activity under feedback control for $\tau \to \infty$ as follows:
\begin{align}
   \mathcal{B}_{\mathrm{\infty}}(\tau)\equiv\tau(\mathfrak{a}+\mathfrak{b}_{c})=\tau\left(\mathfrak{a}+4\mathcal{Z}_{1}+4\mathcal{Z}_{2}\right),
    \label{eq:Bt_long_tau_app}
\end{align}
where $\mathcal{Z}_1$ and $\mathcal{Z}_2$ are defined by
\begin{align}
    \mathcal{Z}_{1}&=-\dblbra{\mathbb{I}}\hat{\mathcal{K}}_{1}(I-\dblket{\rho_{S}^{\mathrm{ss}}}\dblbra{\mathbb{I}})\hat{\mathcal{L}}^{+}(I-\dblket{\rho_{S}^{\mathrm{ss}}}\dblbra{\mathbb{I}})\hat{\mathcal{K}}_{2}\dblket{\rho_{S}^{\mathrm{ss}}},\label{eq:mathcalZ1_def}\\\mathcal{Z}_{2}&=-\dblbra{\mathbb{I}}\hat{\mathcal{K}}_{2}(I-\dblket{\rho_{S}^{\mathrm{ss}}}\dblbra{\mathbb{I}})\hat{\mathcal{L}}^{+}(I-\dblket{\rho_{S}^{\mathrm{ss}}}\dblbra{\mathbb{I}})\hat{\mathcal{K}}_{1}\dblket{\rho_{S}^{\mathrm{ss}}},
    \label{eq:mathcalZ2_def}
\end{align}
where $\hat{\mathcal{L}}$ is the vectorilization of $\mathcal{L}$ and $+$ denotes the  Moore-Penrose pseudoinverse. 
In Eq.~\eqref{eq:mathcalZ2_def}, 
$\hat{\mathcal{K}}_1$ and $\hat{\mathcal{K}}_2$ are vectoralization of the following operators:
\begin{align}
    \mathcal{K}_{1}\rho_{S}&\equiv-iH_{S}\rho_{S}+\frac{1}{2}\sum_{z}\left(e^{\nu_{z}\mathcal{F}}L_{z}\rho_{S}L_{z}^{\dagger}-L_{z}^{\dagger}L_{z}\rho_{S}\right),\label{eq:mathcalK1_def}\\\mathcal{K}_{2}\rho_{S}&\equiv i\rho_{S}H_{S}+\frac{1}{2}\sum_{z}\left(e^{\nu_{z}\mathcal{F}}L_{z}\rho_{S}L_{z}^{\dagger}-\rho_{S}L_{z}^{\dagger}L_{z}\right).\label{eq:mathcalK2_def}
\end{align}